\documentclass{aa}  

\usepackage{graphicx}
\usepackage{txfonts}
\usepackage{ulem}
\usepackage{subfigure}
\usepackage{mathtools}
\usepackage{enumitem}
\usepackage[utf8]{inputenc}
\usepackage{tikz,xcolor,hyperref}

\usepackage{color}

\begin{document}

\title{On Variation of Light Curves and Broad Emission Lines for Periodic QSOs from co-rotating Supermassive binary black holes in elliptical orbits}

\titlerunning{BBHs in elliptical orbits}

\author{Junqiang Ge\inst{1,2},
	Youjun Lu\inst{2,1},
	Changshuo Yan\inst{1,2}
	\and
	Jifeng Liu\inst{1,2,3,4,5}
	}
\authorrunning{Ge et al.}
	
\institute{National Astronomical Observatories, Chinese Academy
           of Sciences, 20A Datun Road, Beijing 100101, China\\
	   \email{jqge@nao.cas.cn}
	\and
	   School of Astronomy and Space Science, 
	   University of Chinese Academy of Sciences, Beijing 100049, China\\
	   \email{luyj@nao.cas.cn}
	\and
	   New Cornerstone Science Laboratory, National Astronomical Observatories, 
	   Chinese Academy of Sciences, Beijing 100101, People's Republic of China\\
	\and
	   Institute for Frontiers in Astronomy and Astrophysics, Beijing Normal University, 
	   Beijing, 102206, People's Republic of China\\
	\and
	   WHU-NAOC Joint Center for Astronomy, Wuhan University,
	   Wuhan, People's Republic of China
           }

\date{Received ; accepted }

\abstract
{Periodic QSOs are considered as candidates of supermassive binary black hole (BBH) systems in galactic centers. Their light curves' periodicity can be interpreted as due to the Doppler boosting caused by the rotation of the two black holes (BHs). Further confirmation of these candidates may require different lines of observational evidences.}
{Assuming the Doopler boosting scenario, in this paper we investigate the (coherent) variations of both broad emission lines (BELs) and continuum light curves for active BBH systems surrounding by a circumbinary broad line region (cBLR) and focus on their dependence on the eccentric orbital configuration.} 
{We calculate the variation of continuum light according to the motion of BBHs on elliptical orbits, with simplified orbital orientation for demonstration, the Doppler enhanced/weakened photoionization of each BLR cloud by the central BBH sources and its variation by assuming a shifted $\Gamma-$ distribution of BLR clouds for a simple BLR geometry, and finally obtain the coherent variation of the continuum and the BELs. 
}
{We find that both the amplitude and variation pattern of the continuum light curves and the evolution of the BEL profiles sensitively depend on the eccentric orbital configuration of BBH systems, especially when the eccentricity is large. If only the secondary BH is active, both the variation amplitudes of continuum light curves and BELs increase with increasing BBH inclination angles and orbital eccentricities, but decrease with increasing BBH mass ratio.
If both BHs are active, the asymmetry in the ionization of BLR clouds at different areas caused by the Doppler boosting effect of the secondary BH is weakened due to that of the primary BH at the opposite direction, which leads to systematically smaller variation amplitudes of both continuum light curves and BELs compared with those in the cases with only secondary BH activated.
} 
{The coherent variations of the BEL profiles with the continuum light for those periodic QSOs provide an important way to confirm the existence of BBHs in their center. Future joint analysis of the light curves and multi-epoch observed BEL profiles for periodic QSOs may lead to the identification of a number of BBH systems. 
}


\maketitle

\section{Introduction}
\label{sec:introduction}

Supermassive binary black hole (BBH) systems are natural products of the hierarchical galaxy mergers in the $\Lambda$ cold dark matter ($\Lambda$CDM) cosmological frame \citep[e.g.,][]{Begelman1980, Yu2002}, since almost every massive galaxy hosts a supermassive black hole (BH) at its centre \citep[e.g.,][]{KR1995,Magorrian1998,KH2013}.
Recently, Pulsar Timing Array (PTA) observations reported the evidence of the stochastic nanohertz gravitational-wave (GW) background with detected Hellings-Downs correlation \citep{HD1983} 
at the $\sim 2- 4 \sigma$ confidence level
\citep[e.g.,][]{Agazie2023,Antoniadis2023,Reardon2023,Xu2023}, which is compatible with the predicted GWs produced by the cosmic BBHs \citep[e.g.,][]{Agazie2023b,Chen2023,Ellis2024}.
Individual sub-pc BBHs with periods of about a year are also the main targets of PTA experiments \citep[e.g.,][]{BS2019} and expected to be detected soon \citep[e.g.,][]{Chen2023}.
Electromagnetic observational evidences of such BBHs were also searched for more than a few decades. Although a large number of candidates were selected according to various observational signatures, but definitive evidence for sub-pc BBHs is still elusive \cite[see a recent review by][]{DC2023}. 

Current BBH candidates at sub-parsec separations are mainly found by indirect signatures, such as double-peaked/asymmetric broad emission lines (BELs)
\citep[e.g.,][]{Eracleous1994,Gaskell1996,Tsalmantza2011,Eracleous2012,Popovic2012,Decarli2013,Ju2013,Shen2013,Liu2014,Guo2019},
periodic optical/UV light curves 
\citep[e.g.,][]{DOrazio2015,Graham2015,Charisi2016,Liu2019,Chen2022}, or other methods \citep[e.g.,][]{Yan2015,Zheng2016}.
However, each type of signature could have alternative model interpretation other than the BBH model, which means only taking one of the above signatures is hard to confirm the existence of a BBH system. For example, the Keplerian rotation of a disk can also produce double-peaked BELs as observed \citep[e.g.,][]{CH1989,Eracleous1997}.
The periodic light curves can also be produced by damped random walk of the QSO flux variation among a large QSO sample \citep[e.g.,][]{Vaughan2016,Liu2019}. Direct resolving and identifying of these BBH candidates may require a spatial resolution of $\sim$ micro arcsec or smaller, which is not achievable by current facilities.

Current numbers of BBH candidates selected from various methods 
are on the order $10^2$, which are mainly obtained by the SDSS QSO spectroscopic survey \citep{Tsalmantza2011, Eracleous2012, Liu2014}, the Catalina Real-time Transient Survey (CRTS), Panoramic Survey Telescope and Rapid Response System (Pan-STARRS), Palomar Transient Factory (PTF), and Zwicky Transient Facility (ZTF) photometric surveys \citep{Graham2015,Charisi2016,Liu2019,Chen2022}, respectively. Confirmation of these BBH candidates may requires both further observations and detailed theoretical modelling, in which observations include continuous photometric monitoring to verify the periodicity \citep[e.g.,][]{Liu2018} and spectroscopic observations to reveal the response of BELs \citep[e.g.,][]{Li2016}, and detailed theoretical modeling is necessary to establish BBH+BLR systems to study how the BELs vary with the co-rotating BBHs \citep[e.g.,][]{Ji2021a, Ji2021b, Songsheng2023}. The combination of photometric/spectroscopic monitoring and theoretical modeling may improve the interpretation to these BBH candidates at subparsec separations \citep[e.g.,][]{DOrazio2015, Song2020, Song2021}.

When applying theoretical models to interpret observed light curves or BEL profiles of QSOs, a simple assumption was made frequently that only the secondary BH is active and the two BHs co-rotating on a circular orbit \citep[e.g.,][]{DOrazio2015, Song2020, Song2021, Ji2021a, Ji2021b, Songsheng2023}. However, both observations and simulations indicate that BBH systems may have high orbital eccentricities \citep[e.g.][]{Valtonen2021,Jiang2022,Lai2023}.
Hydrodynamic simulations also suggest that the accretion mode of BBHs depend on several parameters, including the mass ratio, eccentricity, disk thickness, and kinematic viscosity of the accretion disk(s) \citep[e.g.,][]{Miranda2017, Duffell2020, DD2021, DR2022}. The accretion rate of the secondary BH (with mass $M_2$) tends to be substantially higher than that of the primary BH (with mass $M_1$) when $q_M=M_2/M_1$ is substantially smaller than $1$, and it becomes equal to that of the primary one when $q_M\sim 1$ \citep[e.g.,][]{Farris2014, Kelley2019, Duffell2020, DR2023}.

The on-going and planned photometric and spectroscopic surveys will find a large number of QSOs with periodic light curves and obtain their multi-epoch spectra.
These surveys include the Vera Rubin Observatory’s Legacy Survey of Space and Time \citep[Rubin,][]{Ivezic2019}, WFST \citep{WangT2023}, ZTF \citep{Bellm2019}, DESI \citep{DESIsurvey2016}, the Nancy Grace Roman Space Telescope \citep{Spergel2015}, SDSS-V \citep{Kollmeier2017}, and the Sitian project \citep{Liu2021}. Combining the observed periodic light curves and the variation of the BEL profiles together could provide coherent evidences for the existence of BBHs in these QSOs. Therefore, it is of great importance to explore the dependence of the variation pattern of the periodic light curves and the BEL profiles on the orbital parameters of BBH systems.

In this paper, we focus on investigating the variation pattern of periodic light curves and BELs of active BBH systems with different accretion mode and orbital eccentricities associated with a circumbinary BLR (cBLR). 
This paper is organized as follows.
In Section 2, we describe the basic procedures for establishing the BBH model system. In section 3, we show the main results of light curve and BEL variations of hypothesized BBH systems with different accretion modes and eccentricities. In Section 4, we discuss the uncertainties and applications of the current model. The conclusions are given in Section 5.

\section{Models of BBHs with circumbinary BLR systems}

\begin{figure}
\centering
\includegraphics[angle=0.0,scale=0.47,origin=lb]{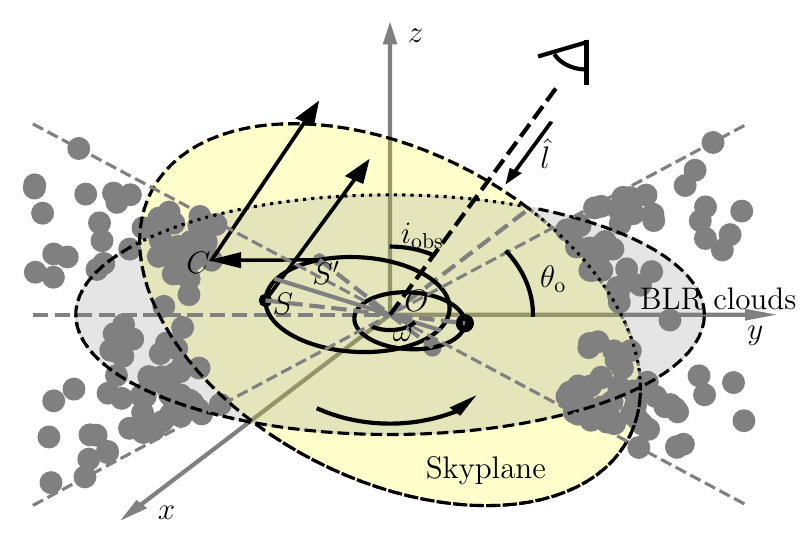}
\caption{Sketch diagram for the configuration of a BBH+cBLR system, with the BBHs and BLR clouds rotating in the same direction. For the BBHs co-rotating in elliptical orbits, the longitude of ascending node is set to be $\Omega=0^\circ$ for simplicity, i.e., the sky plane (yellow) and the BBH orbital plane (grey) share the same $x$-axis in the barycentric coordinate system. The argument of periapsis is denoted as $\omega$.
The observer's line of sight is set in the $y-z$ with an inclination angle of $i_{\rm obs}$ to the BBH orbital plane. The BLR has an opening angle of $\theta_{\rm o}$ and its middle plane is the same as the BBH orbital plane.
}
\label{Fig:ecc_orbits}
\end{figure}
\begin{table}
\caption{
Model parameters for BBH systems co-rotating in elliptical orbits surrounded by a circumbinary BLR.
}
\centering
\begin{tabular}{cc} 
\hline \hline
Parameters & Values \\ \hline
$M_{\bullet\bullet}$ & $10^9M_{\sun}$ \\
$T_{\rm orb}$ & 2 yr \\
$q_M$ & 0.1, 0.5 \\
$q_L$ & 1:0, 1:1 \\
$e$ & 0 (circular orbit), 0.5 \\
$\omega$ & $0^{\circ}$, $90^{\circ}$, $180^{\circ}$, $270^{\circ}$ \\
$\phi$ & 0, $\pi/3$, $2\pi/3$, $\pi$, $4\pi/3$, $5\pi/3$ \\
$i_{\rm obs}$ & $30^{\circ}$, $60^{\circ}$ \\
$\theta_{\rm o}$ & $45^{\circ}$ \\ \hline \\
\end{tabular}
%
\tablefoot{
Rows from top to bottom list the total mass of the BBH system ($M_{\bullet\bullet}$), orbital period of the co-rotating BHs ($T_{\rm orb}$), mass ratio $q_M$ and luminosity ratio $q_L$ of the two active BHs,
eccentricity ($e$) of the BBH orbit, the argument of periapsis ($\omega$), the true anomaly of the secondary BH in one period ($\phi$), viewing angle $i_{\rm obs}$ ($0^\circ$ means face-on), and half opening angle of the BLR ($\theta_{\rm o}$), respectively. The total bolometric luminosity of the BBH system is assumed to be $L_{\rm bol} = 0.1 L_{\rm Edd}^{\bullet\bullet}=1.26\times 10^{37} M_{\bullet\bullet}/M_{\sun}$ erg/s. 
}
\label{tbl:t1}
\end{table}

For the BBH candidates selected from periodic light curves, the typical values of the BBH total mass and orbital period are $M_{\bullet\bullet}\sim 10^9 M_{\sun}$ and $T_{\rm orb}\sim 2$ yr, respectively \citep[e.g.,][]{Graham2015, Charisi2016}. In this work, we hence take BBH systems with these typical mass and period for case studies, by focusing on how the continuum flux and BEL profiles 
are dependent on
mass ratio ($q_M$), 
BH accretion mode (denoted by the UV photoionization luminosity ratio of the two components $q_L$),
orbital eccentricities ($e$), arguments of periapsis ($\omega$), true anomaly ($\phi$), and inclination angles ($i_{\rm obs}$). Here we define 
the luminosity ratio as $q_L=L_{\rm bol}^2/L_{\rm bol}^1$, with $L_{\rm bol}^2$ and $L_{\rm bol}^1$ representing the bolometric luminosities from the accretion disks around the secondary and primary BHs, respectively. Here the bolometric luminosity of the $i$-th (primary/secondary) BH $L_{\rm bol}^i=\epsilon \dot{M}_i c^2$, where $\epsilon$ is the mass-to-energy conversion efficiency, $\dot{M}_i$ is the mass accretion rate of the $i$-th BH, and $c$ is the speed of light \citep[e.g.,][]{YT2002}.
Fig.~\ref{Fig:ecc_orbits} shows the sketch diagram of the BBH+cBLR system.
Here we simplify the system by setting the longitude of ascending node $\Omega=0^{\circ}$ to clarify variations of continuum light curves and BELs that are caused by $e$ and $\omega$.

The detailed model parameters are listed in Table \ref{tbl:t1}. 
The mass ratios of the two BHs $q_M$ are set to be either 0.1 or 0.5, which correspond to remnants by a minor or major merger.
According to numerical simulations, the difference of mass accretion rates $\dot{M}_2$ and $\dot{M}_1$ can be simply represented as a function of $q_M$, e.g., $\dot{M}_2/\dot{M}_1 = 1/(0.1+0.9 q_M)$ \citep[][]{Farris2014, Duffell2020}.
When $q_M=0.1$, $\dot{M}_2$ is $\sim 5$ times higher than $\dot{M}_1$. When $q_M=0.5$, $\dot{M}_2$ is $\sim 2$ times higher than $\dot{M}_1$.
In this work, we simply assume two kinds of accretion modes for BBH systems: 1) Only the secondary BH is active, i.e., $q_L=1:0$, which corresponds to those cases that the secondary BH dominates the flux variations in the Doppler boosting scenario. 2) Both BHs are active with 
$q_L=1:1$, which corresponds to those cases with mass ratio of $q_M\lesssim 1$ where the accretion rates of the two BHs are about the same. Considering that the real observations are more complex than simulations that with simplified parameter input, the analyses of current case study will include both the setup of $(q_M,q_L)=(0.1,1:0)$ and $(0.5,1:1)$ as suggested from simulations, and the cases of $(q_M,q_L)=(0.1,1:1)$ and $(0.5,1:0)$ for comparison.

Considering that the BELs, e.g., $\rm H\alpha$ and $\rm H\beta$, are photoionized by UV photons emitted from the accretion disk. For the BBH+cBLR system, the BBHs and BEL profiles are correlated only if the accretion disk of the active BH could radiate UV photons. Here we simply assume that the photoionization of BLR clouds by the UV photons emitted from the accretion disk of an active BH is effective only if the radius of Roche lobe ($R_{\rm crit}$) is larger than the typical radius of UV continuum emission ($r_{\rm UV}$) (see Appendix \ref{accretion_disk_stab} for detailed analyses). Only with $R_{\rm crit}\gtrsim r_{\rm UV}$ we can study the photoionization of BLR clouds by the central BHs. As shown in Fig. \ref{Fig:BBH_dist_vel}, $e\lesssim 0.6$ is a reasonable criterion for studying the Doppler boosting effect of BBH+cBLR systems with $M_{\bullet\bullet}\sim 10^9 M_{\sun}$ and $T_{\rm orb}\sim 2$ yr. In the case study, we set $e=0$ and $0.5$ for comparing the effects of circular and elliptical orbits of BBHs to the observed continuum and BEL flux variations.

We derive the semi-major axis $a_{\rm BBH}$ of BBH systems with given BBH total mass and orbital period as well as the orbital radii of the two BHs, the azimuthal and radial velocity components $V_{\phi}$ and $V_r$ according to the elliptical motion of the two BHs (see Appendix~\ref{BBH_dynamics}). According to the kinematic motion of the two BHs and the accretion mode settings, we can calculated the Doppler boosted continuum light curves and also the Doppler enhanced/weakened photoionization to BLR clouds. In the simplified BLR model (Appendix \ref{BLR_model}), by assuming the BLR size of the BBH system $R_{\rm BLR}$ follows the same empirical relationship with optical luminosity as the single BH case, we find that $R_{\rm BLR}$ is roughly $\sim 8$ times larger than the separation of BBHs for those candidates selected by periodic light curves. This suggests that the BBH candidates have circumbinary BLRs. By setting a shifted $\Gamma-$ distribution \citep[e.g.,][]{Pancoast2014a} of BLR geometry,
we can calculate the response of BLR clouds to the continuum variation of the central ionizing source (Appendix~\ref{BLR_response}) by considering the following factors: 1) position variation of the two BH components, 2) time-dependent Doppler boosting/weakening effect of the ionizing flux to each BLR cloud, 3) the line emission from each BLR cloud with reverberation response to the central ionizing source, 4) the gravitational redshift of the photons emitted from each BLR cloud, and 5) the Doppler shifts of the photons emitting from each BLR cloud to the observer. By considering all these factors, we finally derive the coherent variation of BEL profiles with the continuum.

In the Doppler boosting scenario, the amplitudes of enhanced/weakened continuum and BEL fluxes depend on many parameters such as those listed in Table \ref{tbl:t1}. For an elliptical orbit, as shown in Fig.~\ref{Fig:ecc_orbits}  by assuming the longitude of ascending node $\Omega=0^\circ$ for simplification, we sample the argument of periapsis $\omega$ ranging from  $0^\circ$ to $270^\circ$ linearly with a step of $90^\circ$ to show the dependence to the variation of the observed continuum light curves on it. And we take six snapshots, i.e., $\phi=0$, $\pi/3$, $2\pi/3$, $\pi$, $4\pi/3$, and $5\pi/3$ to monitor the phase evolution of BEL profiles modulated by the orbital motion of BBHs. The observer is set to have a viewing angle of either $i_{\rm obs}=30^\circ$ (close to face-on) or $60^\circ$ (close to edge-on). The half opening angle of the BLR is set as $45^\circ$ \citep[flattened disk geometry,][]{Gravity2018}, 
with all the BLR clouds rotating in circular orbits. As to the rotating direction of the two BHs and BLR clouds, all of them are assumed to be counterclockwise rotating.

\section{Results}

\subsection{Variation of continuum light curves from BBHs in elliptical orbits}\label{Sec:light_curves}

\begin{figure*}
\centering
\includegraphics[angle=0.0,scale=0.95,origin=lb]{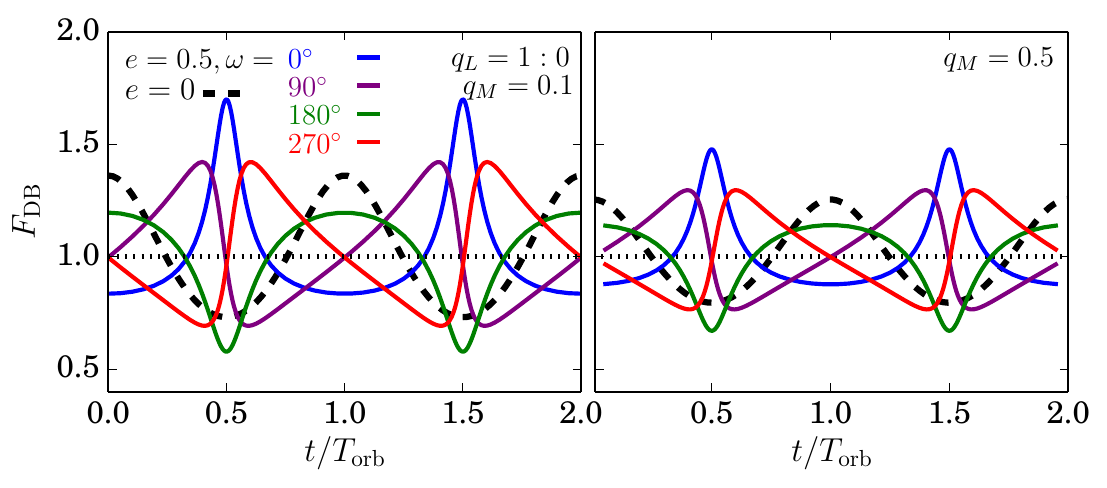}
\caption{Comparison of continuum light curves emitted by BBH systems when only the secondary BH is active with mass ratio $q_M=0.1$ (left panel) and $0.5$ (right panel) at $i_{\rm obs}=60^\circ$. The black dashed line shows the light curve from the system with circular orbit ($e=0$), the other four lines coloured from blue to red show those resulted from elliptical orbits ($e=0.5$), with four different position angles of their major axes $\omega$ varying from $0^\circ$ to $270^\circ$. The horizontal dotted line shows the light curve without the Doppler Boosting effect.}
\label{Fig:conti_light_curve}
\end{figure*}
\begin{figure*}
\centering
\includegraphics[angle=0.0,scale=0.95,origin=lb]{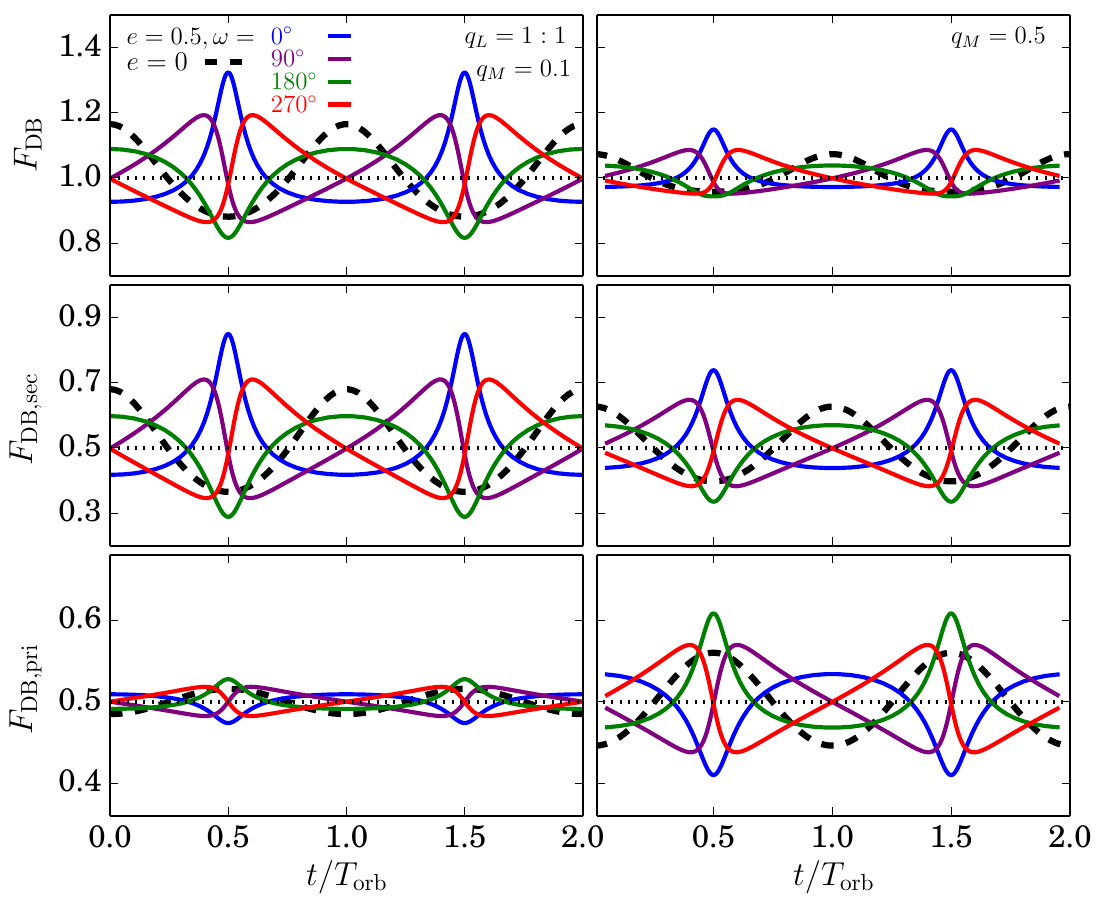}
\caption{Comparison of continuum light curves emitted by BBH systems when both the two BHs are active, with $q_L=1:1$, $i_{\rm obs}=60^\circ$, $q_M=0.1$ (left column) and $0.5$ (right column). The top panel shows the observed continuum light curve contributed by Doppler Boosting effect from the two co-rotating black holes. The Doppler Boosted/weakened flux modulated by the secondary and the primary BHs are shown in the middle and bottom panels, respectively. Lines in different shapes and colours are the same as shown in Fig. \ref{Fig:conti_light_curve}.}
\label{Fig:conti_light_curve_2active}
\end{figure*}

For an BBH system co-rotating in elliptical orbit surrounded by a circumbinary BLR, the continuum light curve received by the observer depends on not only $M_{\bullet\bullet}$, $T_{\rm orb}$, $q_M$, and $q_L$, but also $e$ and $\omega$, compared to the case of circular orbit. 

\subsubsection{$q_L=1:0$}

In the case of $q_L=1:0$, Fig. \ref{Fig:conti_light_curve} compares the continuum light curves obtained for cases with circular ($e=0$, black dashed line) and elliptical orbits ($e=0.5$ and $\omega=0^\circ$ to $270^\circ$, colored from blue to red) with mass ratio $q_M=0.1$ (left panel) and 0.5 (right panel). When only the secondary BH is active, its rotational velocity increases with decreasing $q_M$, which consequently leads to smaller variation amplitudes of the light curves for the system with $q_M=0.5$ than those with $q_M=0.1$.

For a counter-clockwise rotating BBH+cBLR system, the evolution trend of the observed continuum light curve is mainly determined by the increase/decrease of the projected velocities to the observer. As shown in Fig. \ref{Fig:ecc_orbits}, for the cases of $\omega=0^\circ$ and $180^\circ$, the tangential lines at the pericenter are parallel to the observer's plane (Y-Z plane for the current setup), which result symmetric light curves at the first half and second half in each period as that of the circular orbit, but with different Doppler enhanced/weakened amplitudes and time durations. In addition to the counter-clockwise rotating direction of the secondary BH, one can observe a continuum light curve with 
the strongest Doppler enhanced and the smallest Doppler weakened amplitudes at $\omega=0^\circ$, and a light curve with the smallest Doppler enhanced and the largest Doppler weakened amplitudes at $\omega=180^\circ$.
For other cases, i.e., $\omega \neq 0^\circ$ or $180^\circ$, the observed continuum light curves are asymmetric. The shape of a continuum light curve is determined by the projected velocities of the secondary BH at the pericenter. When $0^\circ<\omega< 180^\circ$, the light curves show slow increase and rapid decrease variation pattern (e.g., $\omega= 90^\circ$ in Fig. \ref{Fig:conti_light_curve}). When $180^\circ<\omega< 360^\circ$, the light curves show rapid increase but slow decrease variation pattern (e.g., $\omega= 270^\circ$ in Fig. \ref{Fig:conti_light_curve}).

In the case of $q_M=0.5$ (right panel of Fig. \ref{Fig:conti_light_curve}), the rotational velocities at each phase are systematically smaller than that of $q_M=0.1$ (left panel of Fig. \ref{Fig:conti_light_curve}), the corresponding Doppler enhanced/weakened amplitudes are hence also systematically smaller, but keeping the variation patterns the same as that of $q_M=0.1$. 

\subsubsection{$q_L=1:1$}

When both the two BHs are active, the amplitude of the continuum light curves depends on the luminosity ratio of the two BHs. The secondary BH has higher rotational velocity than the primary one, and hence contribute more to the Doppler enhanced/weakened amplitudes. As shown in the left column of Fig. \ref{Fig:conti_light_curve_2active}, for $q_L=1:1$ and $q_M=0.1$, the emitted continuum light curve by the secondary BH is the same as that shown in Fig. \ref{Fig:conti_light_curve}, but with the assumed intrinsic flux decreased from 1 to 0.5. The amplitude of continuum light curve emitted by the primary BH is $\sim 10$ times weaker than the secondary one, and the shape of the light curve is opposite to that of the secondary one due to the momentum conservation. With the mass ratio $q_M$ increasing to $0.5$ (right column of Fig. \ref{Fig:conti_light_curve_2active}), the Doppler boosted amplitude of the secondary BH is systematically smaller than that of $q_M=0.1$. On the other hand, the rotational velocity of the primary BH increases, which causes a higher amplitude of the light curve than the case of $q_M=0.1$. The increasing mass ratio $q_M$ indicates a decreasing (increasing) amplitude of light curves by the secondary (primary) BH, which means that at any time, the enhanced (weakened) flux modulated by the secondary BH always correspond to the weakened (enhanced) flux by the primary BH. 
This makes the observed continuum light curves having smaller amplitudes than the case of $q_L=1:0$. With increasing $q_M$, resolving the periodicity of these light curves hence require higher quality of photometric observations.

With the interpretation on the behaviours of the light curve variations from BBH systems, we then investigate how the response of the circumbinary BLR vary at different cases.

\subsection{The BEL profile variations for different BBH systems in circular orbits} \label{Sec:BBH_circ}
\begin{figure*}
\centering
\includegraphics[angle=0.0,scale=0.65,origin=lb]{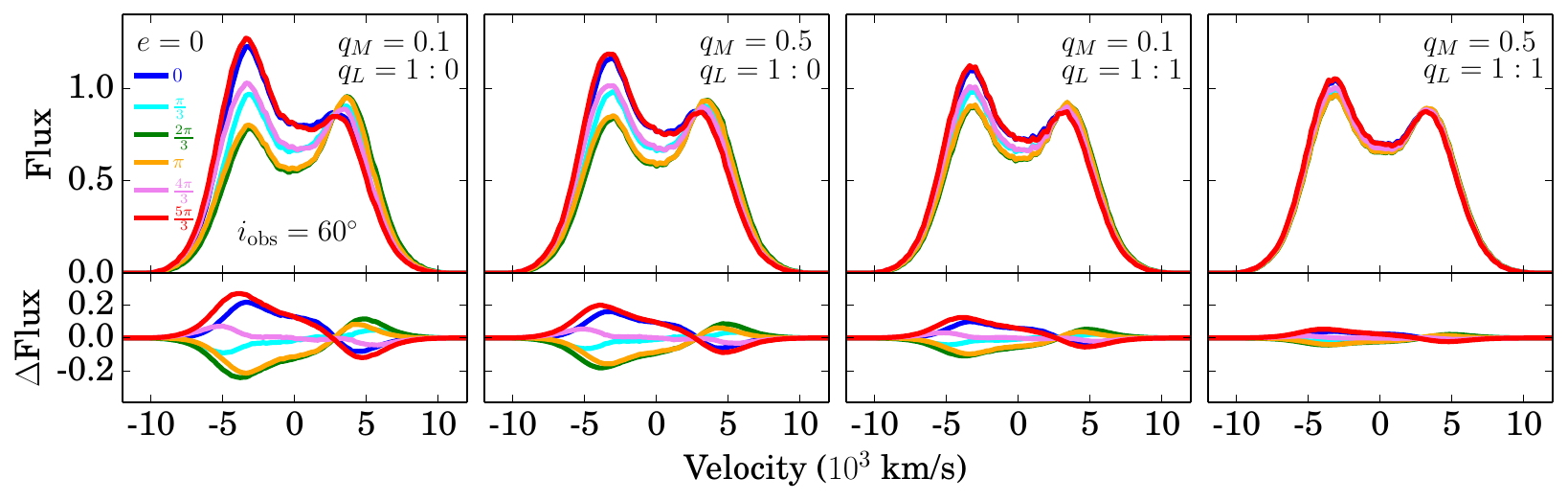}
\caption{BEL profile variations in a single orbital period for BBHs co-rotating in circular orbits with $i_{\rm obs}=60^\circ$. The left two columns show BBH systems when only the secondary BH is active, i.e., $q_L=1:0$, but with $q_M=0.1$ (left column) and $0.5$ (middle-left column). The right two columns show profile variations when both the two BBHs are active with $q_L=1:1$, in the case of $q_M=0.1$ (middle-right column) and $0.5$ (right column). In each column, the top panel shows BEL profiles of the six phases (0 to $5\pi/3$) in one period labelled in blue to red colours, and the bottom panel presents the relative flux variation of BEL profiles compared to the mean profile obtained from that of the six phases.}
\label{Fig:circ_orbits}
\end{figure*}

To clarify how the elliptical orbits of BBHs can affect the response of BLR clouds, we first study the variation pattern of profile shapes and amplitudes with different mass and luminosity ratios of BBHs co-rotating in circular orbit. Fig. \ref{Fig:circ_orbits} shows the profile variations in one period at six phases ($\phi=0$, $\pi/3$, $2\pi/3$, $\pi$, $4\pi/3$, and $5\pi/3$) for the four models, i.e., $(q_M, q_L)=$ (0.1, 1:0), (0.5, 1:0), (0.1, 1:1), and (0.5, 1:1), respectively. For each model, we show the relative flux variations, which are relative to the mean profile that obtained from the six phases, in each corresponding bottom panel. 
In the case of BBHs co-rotating in the same direction with the BLR clouds, given an observer with $i_{\rm obs}>0$, those blue-shifted (moving toward the observer) clouds contribute higher flux variations than those red-shifted BLR clouds due to the time-delay effect, which are characterized by the relative flux variations at each bottom panel of Fig. \ref{Fig:circ_orbits}.

When only the secondary BH is active ($q_L=1:0$), the amplitude of the light curve for $q_M=0.5$ is slightly smaller than that for $q_M=0.1$ (Fig. \ref{Fig:conti_light_curve}), the BEL profile variations for $q_M=0.1$ (left column of Fig. \ref{Fig:circ_orbits}) and $0.5$ (left-second column of Fig. \ref{Fig:circ_orbits}) are quite similar and the flux variations caused by the Doppler boosting effect for the model of $q_M=0.5$ is also slightly smaller than that of $q_M=0.1$. 

The amplitudes of BEL profile variations decrease significantly with increasing $q_M$ for the case of $q_L=1:1$. As shown in the right two columns of Fig. \ref{Fig:circ_orbits}, the amplitude of flux variations for $q_M=0.5$ is substantially smaller than that for $q_M=0.1$. This dramatic decrease is not due to the Doppler boosting effect by the secondary BH, but is dominated by the ionizing photons emitted from the accretion disk of the primary BH. Although the profile variation caused by the secondary BH has slight difference between $q_M=0.1$ and $0.5$, the difference of rotating velocities and hence Doppler boosting effect of their primary BHs for the two models increases dramatically from the cases with $q_M=0.1$ to those with $q_M=0.5$ (see the bottom panels of Fig. \ref{Fig:conti_light_curve_2active}). 

\subsection{The BEL profile variations for different BBH systems in elliptical orbits -- dependence on $\omega$ and $\phi$}

\begin{figure*}
\centering
\includegraphics[angle=0.0,scale=0.6,origin=lb]{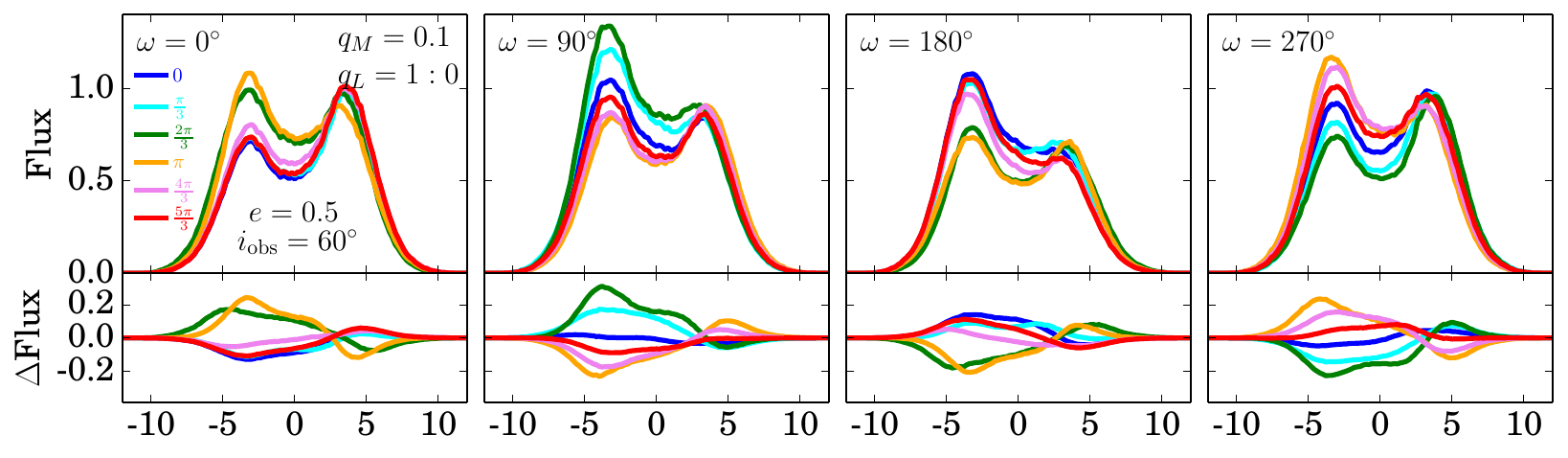}\\
\includegraphics[angle=0.0,scale=0.6,origin=lb]{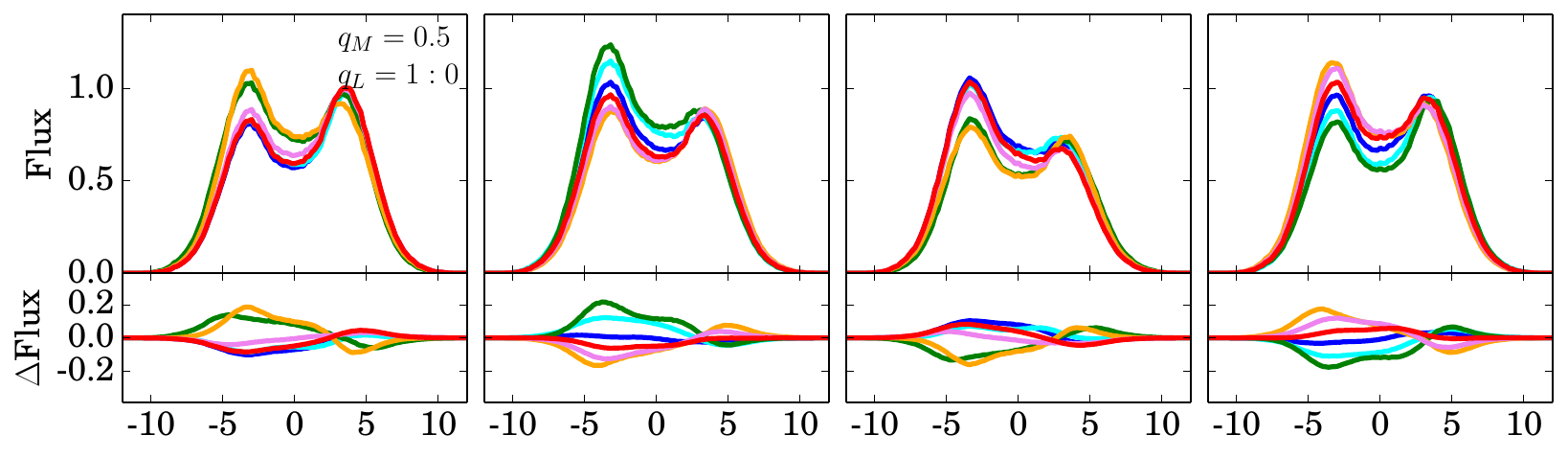}\\
\includegraphics[angle=0.0,scale=0.6,origin=lb]{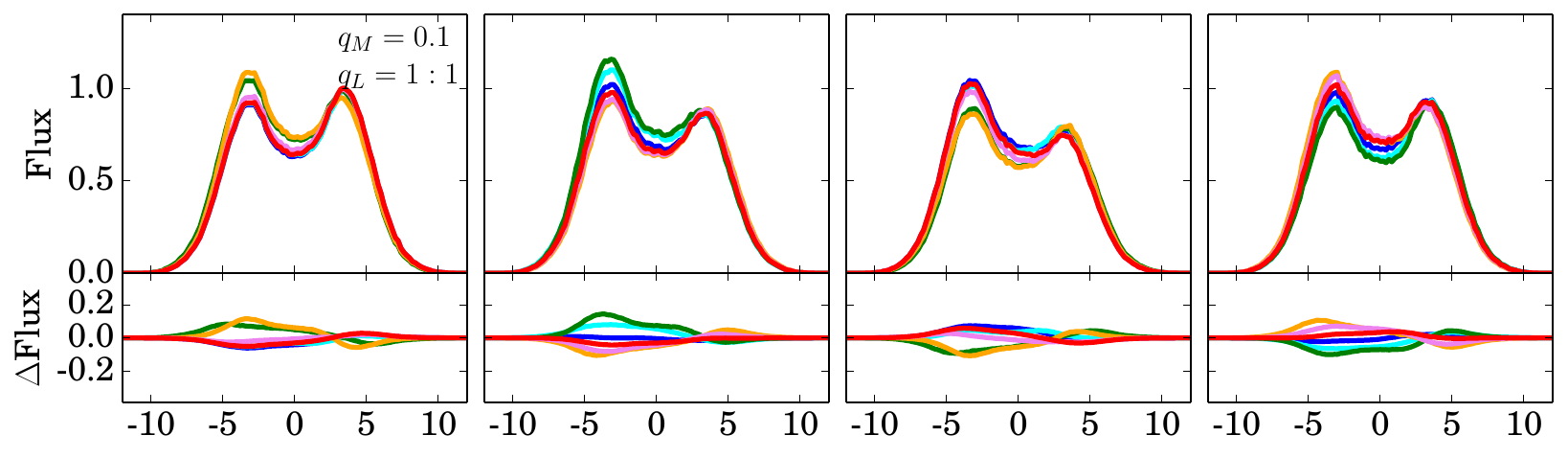}\\
\includegraphics[angle=0.0,scale=0.6,origin=lb]{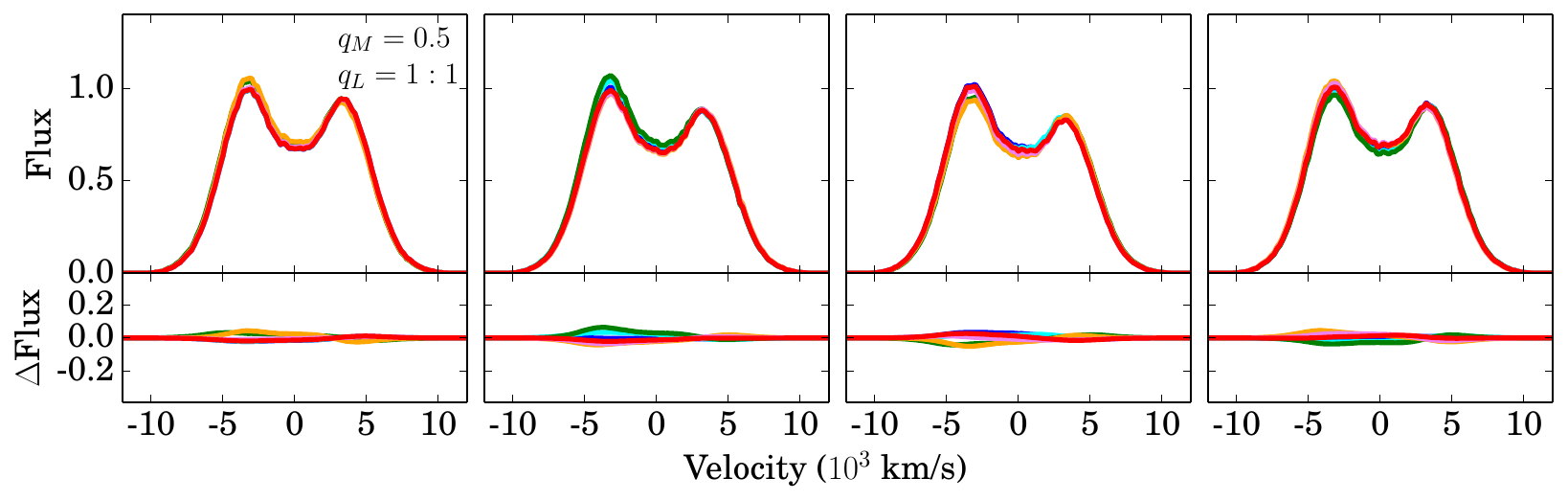}
\caption{Comparison of continuum light curves emitted by BBH systems when only the secondary BH is active (top two rows) and both the two BH are active (bottom two rows) for $e=0.5$ and $i_{\rm obs}=60^\circ$. In each row, the top panels show the profiles at six phases ($0$ to $5\pi/3$) in one period, and the relative flux variations are shown in the corresponding bottom panels as that in Fig. \ref{Fig:circ_orbits}. Columns from left to right show the cases with $\omega=0^\circ, 90^\circ, 180^\circ$, and $270^\circ$, which reflects the profile variations caused by different eccentricity vectors of the elliptical orbits observed at a fixed viewing angle. }
\label{Fig:secBH_both_iobs60}
\end{figure*}
\begin{figure*}
\centering
\includegraphics[angle=0.0,scale=0.6,origin=lb]{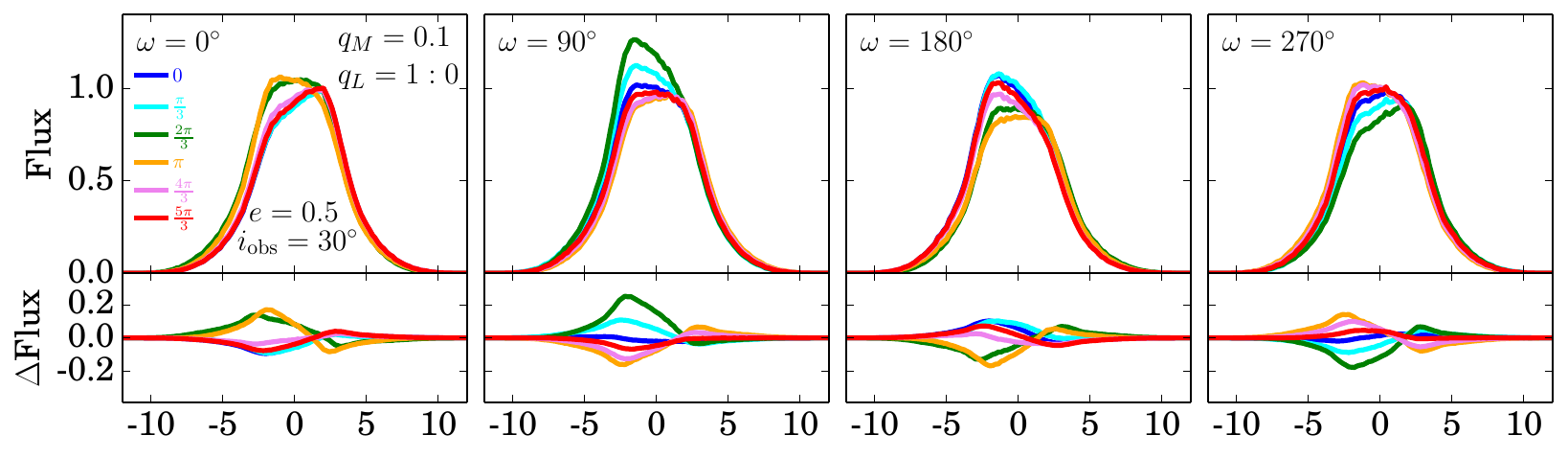}\\
\includegraphics[angle=0.0,scale=0.6,origin=lb]{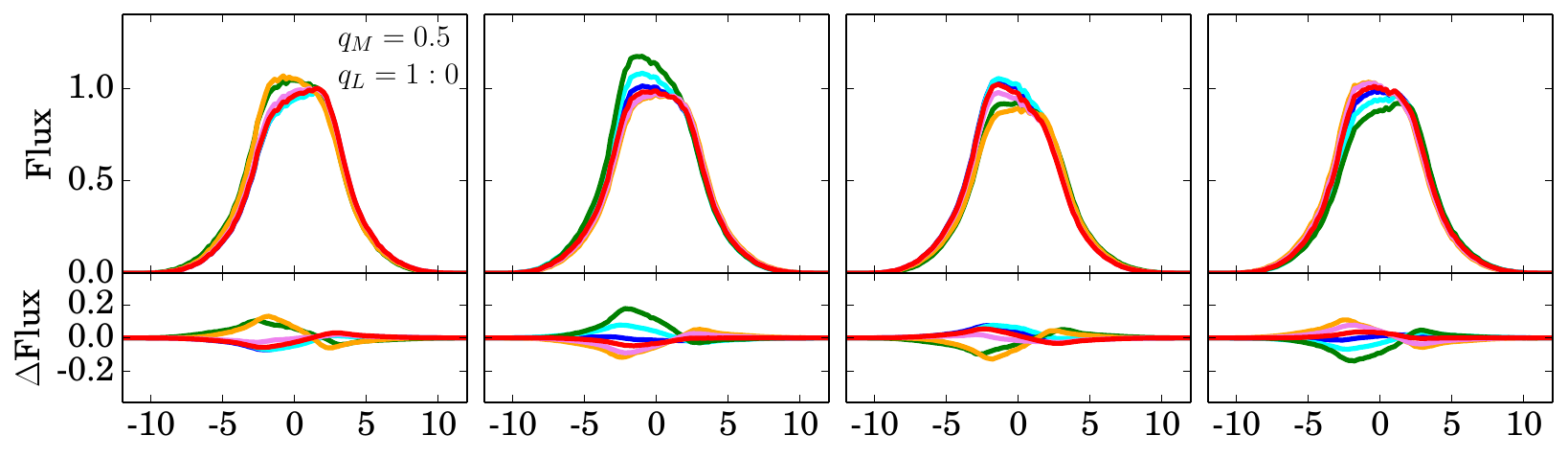}\\
\includegraphics[angle=0.0,scale=0.6,origin=lb]{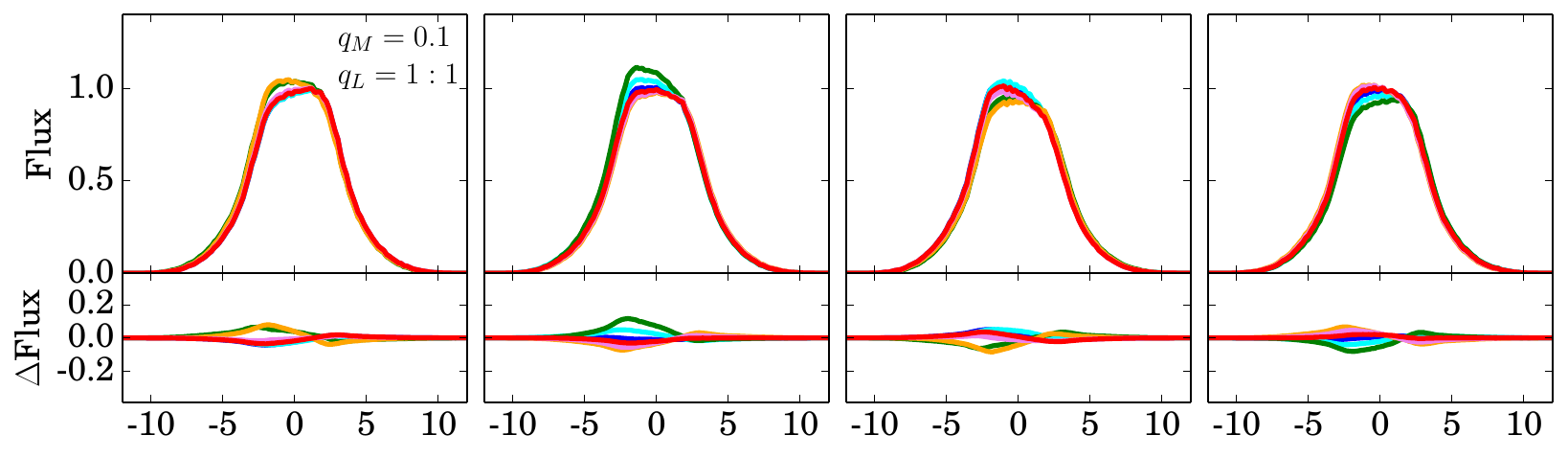}\\
\includegraphics[angle=0.0,scale=0.6,origin=lb]{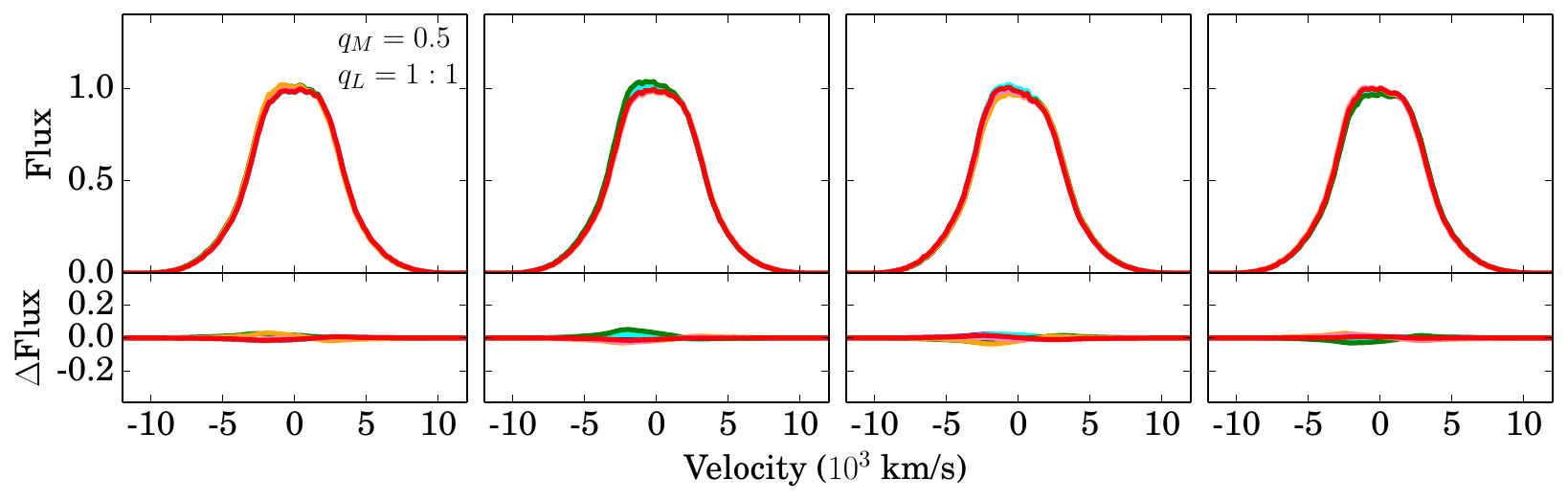}
\caption{Legends are the same as those for Fig. \ref{Fig:secBH_both_iobs60}, except that $i_{\rm obs}=30^\circ$.}
\label{Fig:secBH_both_iobs30}
\end{figure*}

When the two BHs co-rotating in elliptical orbits, the two orbital parameters $e$ and $\omega$ would make more complex profile variations compared to that in circular orbits. For a BBH-cBLR system with fixed $M_{\bullet\bullet}$ and $T_{\rm orb}$, the velocity difference between the apocenter and pericenter increases with increasing eccentricity. Given a viewing angle $i_{\rm obs}$, the variation patterns of profile shapes and fluxes of observed BELs are mainly dominated by $e$ and $\omega$. Different $\omega$, i.e., different projected velocities to the observer and different area of BLR clouds enhanced by the central BH, would make the observed BEL profiles having varied shapes at a certain phase. While the increasing $e$, i.e., rapidly increased velocities at the pericenter (Fig. \ref{Fig:BBH_dist_vel}), would increase the difference of profile shapes among different phases. 

Figs. \ref{Fig:secBH_both_iobs60} and \ref{Fig:secBH_both_iobs30} show the resulting BEL profiles from the Doppler boosting hypothesis of four BBH-BLR systems viewed at $i_{\rm obs}=60^\circ$ and $30^\circ$, respectively. In these two figures, left to right columns show the cases with $\omega=0^\circ$, $90^\circ$, $180^\circ$, and $270^\circ$, respectively. In each panel, phase $0$ means that the counterclockwise rotated secondary BH is located at the pericenter, and phase $\pi$ corresponds to that rotated to the apocenter.

In Fig. \ref{Fig:secBH_both_iobs60}, when the BBH-BLR system is close to edge-on viewed ($i_{\rm obs}=60^\circ$), double-peaked profiles appear for different $\omega$ and phases. As analyzed in Sections \ref{Sec:light_curves} and \ref{Sec:BBH_circ}, the system with ($q_M,q_L$) = (0.1, 1:0) has the highest variation amplitude of light curve and profile variations than the other three systems. The amplitudes of BEL profile variations in one period for elliptic BBH systems are different from those circular ones (Fig. \ref{Fig:circ_orbits}). In the case of $e=0.5$, the BBHs at phases from $\phi=\pi/2$ to $3\pi/2$ that cross the pericenter have shorter rotating timescale and higher velocities than that from $3\pi/2$ to $\pi/2$. For the case that all the BLR clouds are rotating in circular orbits, given the fixed inclination angle of an observer, the BEL profile at a certain phase would not vary. However, for BBHs in elliptical orbit, the varying $\omega$ can cause profile variation at a certain phase. 

With the argument of periapsis $\omega$ varying from $0^\circ$ to $360^\circ$ (left to right columns of Fig. \ref{Fig:secBH_both_iobs60}), BEL profiles have large variations at a given phase, as the positions of each of the two BH components relative to the observer at that phase are different for cases with different $\omega$.
Co-adding with the time-delay effect, the envelope of BEL profile variations in each panel are hence different for those cases with different $\omega$. For example, at $\phi=2\pi/3$ (green line), the double peaked profile has similar fluxes at the blue and red peaks at $\omega=0^\circ$. When $\omega$ changed to $90^\circ$, the blue peak has higher flux than the red peak. At $\omega=180^\circ$, the total flux of the profile increases with the red and blue peak fluxes become similar again. While for $\omega=270^\circ$, the flux of the red peak is significantly higher than that of the blue peak. Therefore, in the elliptic BBH system, two periodic signitures appear. The first one is periodic profile variation in one period, e.g., $\phi=0$ to $5\pi/3$ shown in each panel, which is dominated by rotation of the central BHs. The second one is the periodic profile variation in a certain phase, e.g., $\omega=0^\circ$ to $\sim 270^\circ$ at $\phi=2\pi/3$, which is caused by $\omega$. In the case that the BBH systems have no orbital precession, or the timescale of the precession is hugely longer than the rotating timescale, one can only see the periodic BEL profile variation in one period.

When observing the BBH+cBLR system at a viewing angle close to face-on, e.g., $i_{\rm obs}=30^\circ$ in Fig. \ref{Fig:secBH_both_iobs30}, the BEL profiles only present Gauss-like/asymmetric shapes, and the amplitudes of flux variation decrease significantly for all the four model systems compared to that observed by $i_{\rm obs}=60^\circ$. However, the relative flux variation trends are still similar to that shown in Fig. \ref{Fig:secBH_both_iobs60}. 

\begin{figure*}
\centering
\includegraphics[angle=0.0,scale=0.47,origin=lb]{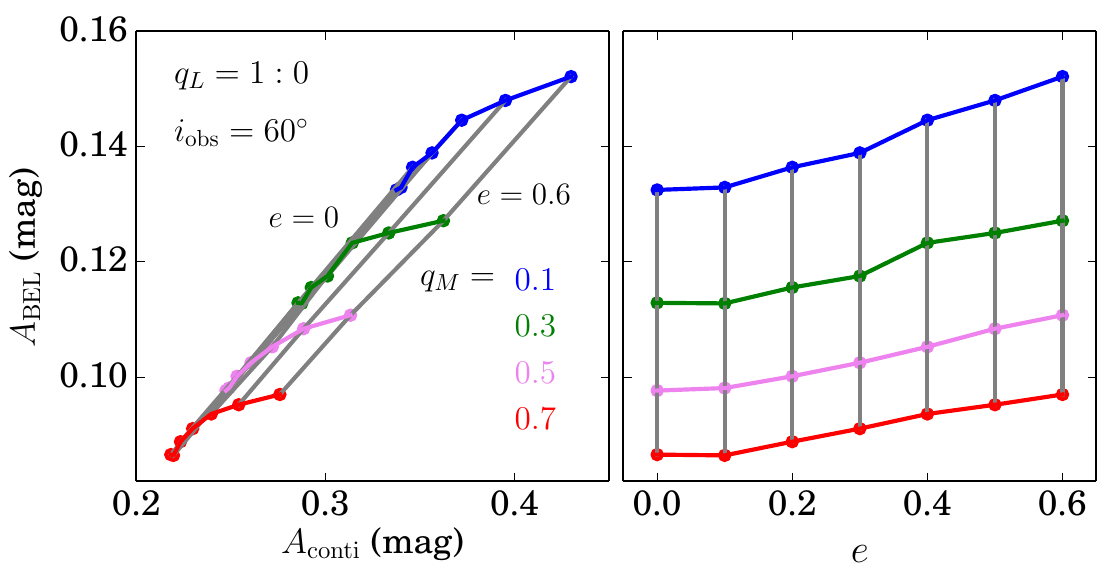}
\includegraphics[angle=0.0,scale=0.47,origin=lb]{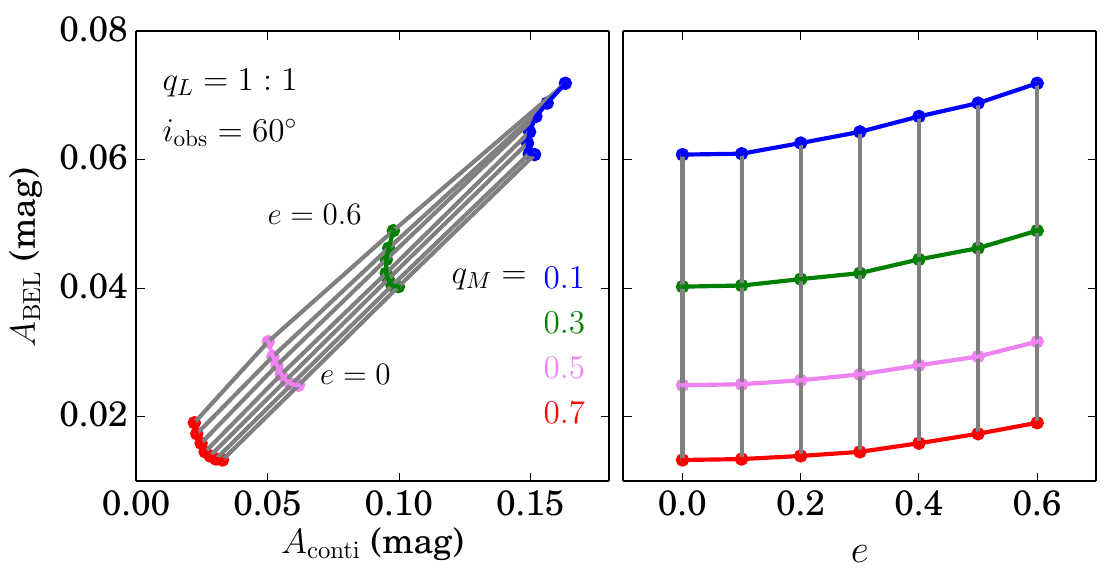}
\caption{Variation of BEL amplitudes ($A_{\rm BEL}$) with different continuum amplitudes ($A_{\rm conti}$) and orbital eccentricities ($e$) of BBHs with $M_{\bullet\bullet}=10^9M_{\sun}$ at $\omega=0^\circ$. The left two columns show $A_{\rm BEL}$ as a function of $A_{\rm conti}$ (left panel) and $e$ (middle-left panel) for the case that only the secondary BH is active ($q_L=1:0$), and the right two columns show $A_{\rm BEL}$ vs. $A_{\rm conti}$ (middle-right panel) and $A_{\rm BEL}$ vs. $e$ (right panel) for the case that both the two BHs are active with $q_L=1:1$. In each panel, points connected by lines in the same color show results of a fixed mass ratio, and the four different mass ratios $q_M=0.1$, $0.3$, $0.5$, and $0.7$ are colored in blue, green, violet, and red, respectively. Each four points with different $q_M$ but the same $e$ are linked by grey lines.}
\label{Fig:secBH_both_conti_BEL_amplitude}
\end{figure*}
\begin{figure*}
\centering
\includegraphics[angle=0.0,scale=0.47,origin=lb]{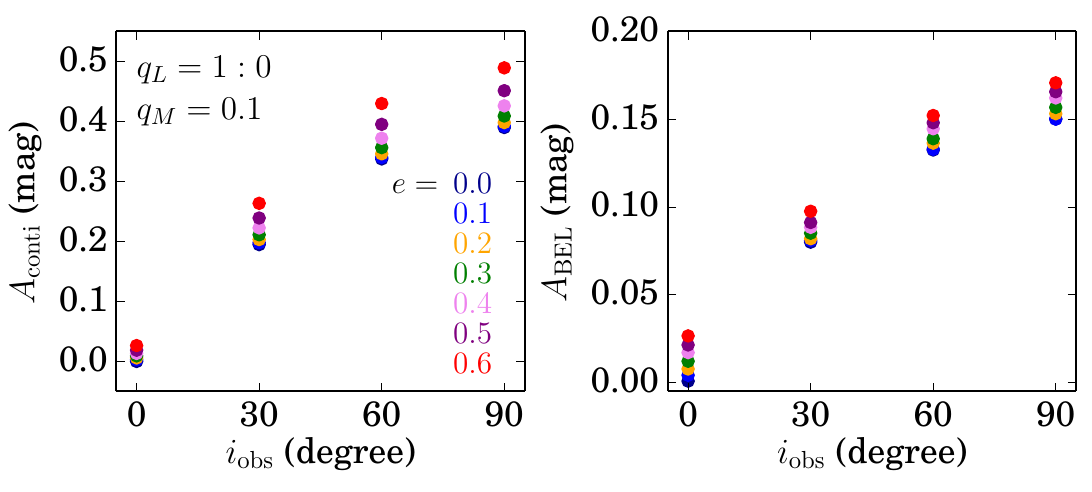}
\includegraphics[angle=0.0,scale=0.47,origin=lb]{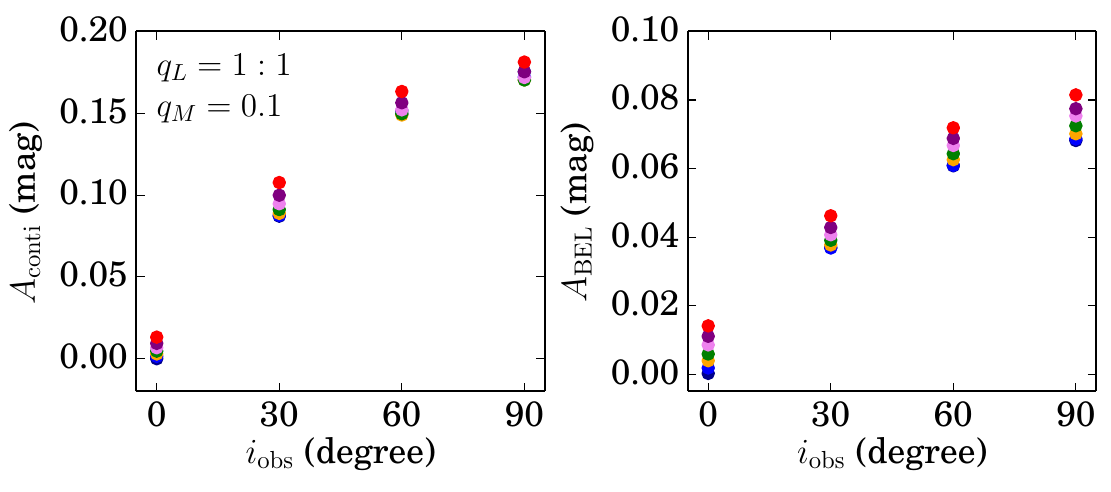}
\caption{Variation of continuum ($A_{\rm conti}$) and BEL amplitudes ($A_{\rm BEL}$) with different inclination angles for BBHs with $M_{\bullet\bullet}=10^9M_{\sun}$, $\omega=0^\circ$, and $e=0.0$ to $0.6$. The left two panels show the case that only the secondary BH is active ($q_L=1:0$), and the right two panels show the case that both the two BHs are active with $q_L=1:1$. In each panel, points colored from dark blue to red represent the orbital eccentricity of BBHs $e=0.0$ to $0.6$, respectively.}
\label{Fig:secBH_both_conti_BEL_diffiobs}
\end{figure*}

\subsection{Varied amplitudes of BELs for different continuum light curves and orbital eccentricities of BBHs}

With the interpretation on the evolution trend of BEL profiles for different $\omega$, we then explore how the amplitude of BEL profiles, i.e., $A_{\rm BEL}=1.25(\log L_{\rm BEL}^{\rm max}-\log L_{\rm BEL}^{\rm min})$, vary with the intrinsic continuum emission by the central ionization sources in different amplitudes, i.e., $A_{\rm conti}=1.25(\log L_{\rm conti}^{\rm max}-\log L_{\rm conti}^{\rm min})$, mass ratio ($q_M$), and orbital eccentricity ($e$). 

When only the secondary BH is active, as shown in the left two panels of Fig. \ref{Fig:secBH_both_conti_BEL_amplitude}, at fixed $e$, higher $q_M$ means lower velocity of the secondary BH, and hence lower $A_{\rm BEL}$ and $A_{\rm conti}$. For the case of $M_{\bullet\bullet}=10^9M_{\sun}$, when the eccentricity $e$ increases from $0$ to $0.6$, the rotating velocity at the pericenter for the secondary BH increase, which corresponds to both increased $A_{\rm conti}$ and $A_{\rm BEL}$. For $A_{\rm BEL}$, its increasing trend slows down with increasing $e$. 
Since the variation of $A_{\rm BEL}$ is caused by two competitive factors: (1) the increasing rotation velocity caused by increasing $e$ at phases $\phi = 3\pi/2$ to $\pi/2$ (at the pericenter $\phi=0$) could contribute stronger Doppler enhancement to the ionization of BLR clouds, (2) the decreasing time duration at phases $\phi = 3\pi/2$ to $\pi/2$ would decrease the fraction of BLR clouds with stronger Doppler enhancement. 

For the case that both the two BHs are active with $q_L=1:1$ (the right two panels of Fig. \ref{Fig:secBH_both_conti_BEL_amplitude}), the response of BLR clouds to the central ionization source is affected by both the secondary and primary BHs. The ionization behaviour of the secondary BH is the same as that of $q_L=1:0$, while the behaviour of the primary BHs is opposite to that of the secondary BH. The differences between flux amplitudes and variation trends shown in the left and right two panels are caused by the activity of the primary BH. The intensity ratio of the two beams depends on both the luminosity ratio $q_L$ and the mass ratio $q_M$. Higher luminosity fraction taken by the primary BH or higher mass ratio of the two BHs indicate enhanced ionization of BLR clouds by the primary BH, which appear in the Doppler weakened regions that modulated by the secondary BH. Therefore, the total fluxes of BELs in the case of $q_L=1:1$ increase and $A_{\rm BEL}$ decreases compared to the case of $q_L=1:0$. This explains why $A_{\rm BEL}$ of $q_L=1:1$ is systematically smaller than that of $q_L=1:0$. 

As shown in the middle-right panel of Fig. \ref{Fig:secBH_both_conti_BEL_amplitude}, the variation of $A_{\rm conti}$ with $e$ is different for increasing $q_M$. At $q_M=0.1$, $A_{\rm conti}$ first decreases and then increases with increasing $e$. At $q_M\gtrsim 0.3$, $A_{\rm conti}$ decreases monotonously with $e$ at larger $q_M$. The value of $A_{\rm conti}$ is determined by the differential amplitude between the primary and secondary BHs (Fig. \ref{Fig:conti_light_curve_2active}) as a function of $e$, since the velocities of the two BHs increase with increasing $e$ (see $V_{\rm peri}^{\rm sec}$ vs. $e$ in the right panel of Fig. \ref{Fig:BBH_dist_vel}). The varying trends of $A_{\rm conti}$ vs. $e$ at different $q_M$ reflect the competitive contribution of continuum emission by the two BHs with the Doppler boosting factor $D^5$ (Eq. \ref{eq:f_BLR}). 

Although the activity of the primary BH decreases both the observed $A_{\rm conti}$ and $A_{\rm BEL}$ (the left and middle right panels of Fig. \ref{Fig:secBH_both_conti_BEL_amplitude}), the BEL profiles still have an increasing $A_{\rm BEL}$ with increasing $e$ (the right panel of Fig. \ref{Fig:secBH_both_conti_BEL_amplitude}), since the response of BLRs to the central ionizing sources is still dominated by the secondary BH.

For $q_L$ varying from $1:0$ to $0:1$, the increasing luminosity fraction of the primary BH correspond to decreasing $A_{\rm conti}$ and $A_{\rm BEL}$ when the secondary BH dominating the Doppler Boosting effect, then the variation trends and phases of the continuum light curve and BEL profiles become inverse when the primary BH dominating the Doppler Boosting effect.

When changing the viewing angle $i_{\rm obs}$ from $0^\circ$ (face-on) to $90^\circ$ (edge-on), similar variation pattern of $A_{\rm conti}$ and $A_{\rm BEL}$ with $e$ appear at different $i_{\rm obs}$. With increasing $e$, $A_{\rm conti}$ and $A_{\rm BEL}$ also increase. On the other hand, the values of $A_{\rm conti}$ and $A_{\rm BEL}$ also increase with increasing $i_{\rm obs}$ for both $q_L=1:0$ and $1:1$ cases (Fig. \ref{Fig:secBH_both_conti_BEL_diffiobs}), which means that the periodic variations of continuum and BEL fluxes are more detectable with larger inclination angles.

In this paper, as a case study, we focus on the continuum and BEL profile variations of the BBH+cBLR system by assuming $\Omega=0^\circ$, i.e., $\cos \Omega=1$, for simplification. For those observed periodic QSOs with arbitrary $\Omega\neq 0^\circ$, if all other parameters are fixed, increasing $\Omega$ means decreased amplitudes of both $A_{\rm conti}$ and $A_{\rm BEL}$. However, their coherent variation behaviours keep the same as that analyzed above.

\section{Discussion}

We have investigated the coherent variations of light curves and BEL profiles for co-rotating BBH systems with different orbital eccentricities, by assuming that either only the secondary is active or both the two BHs are active with equal luminosity. However, the observed cases should be more complex. We hence discuss some complexity in the BBH+cBLR modelling and observational strategy on how to identify BBH systems as follows.

\subsection{Different BLR geometries}

To explain the observed BEL profiles that varying from single peaked to asymmetric or double-peaked shapes, the inclination angle could be changed from face-on to edge-on viewed, and the BLR may also be changed from elliptical \citep[e.g.,][]{Shen2010} to disk-like \citep[e.g.,][]{Eracleous1994} geometries.  For the observed line asymmetry, the disc-like BLRs with clouds in elliptical orbits are preferred for modelling \citep[e.g.,][]{Kovacevic2020}. Some complex BLR structures have also been investigated, e.g., BLRs in thin/thick disk with inflow/outflow substructures \citep[e.g.,][]{Wang2018}, or that with two flattened disks perpendicular to each other \citep[e.g.,][]{Ji2021b}. If considering co-planar BBH+cBLR systems with the two BHs co-rotating in circular orbit, the BLRs varying from spherical to thin disk indicate increasingly sensitive response of BLR clouds to the central source \citep[e.g.,][]{Ji2021a}.

\subsection{Orbital decay of BBHs}

The orbital evolution of two BHs at different semi-major axes is controlled by different decay mechanisms, e.g., from disk-dominated viscous evolution at large separations, to secondary-dominated viscous evolution stage (the distance of BBHs at the pericenter $D_{\rm peri}\lesssim 10^4 R_{\rm Sch}$ for $q_M\sim 0.1$), and then gravitational wave (GW) dominated evolution stage ($D_{\rm peri}\lesssim 500 R_{\rm Sch}$), and finally gas-disk decoupled stage at $D_{\rm peri}\lesssim 100R_{\rm Sch}$, where $R_{\rm Sch}$ is the Schwarzschild radius corresponding to the total mass of BBHs \citep[e.g.,][]{Haiman2009}. 

For BBHs dominated by the GW-driven orbital decay mechanism, the GW decay timescale is
\begin{equation}\label{eq:tgw}
t_{\rm GW} = (8.2\times 10^3 {\rm yr}) \left[\frac{q}{(1+q)^2}\right]^{-1} \left(\frac{M_{\bullet\bullet}}{10^9M_{\odot}}\right)^{-5/3} \left(\frac{T_{\rm orb}}{\rm 2yr}\right)^{8/3} F(e),
\end{equation}
where $F(e)$ is the enhancement factor caused by orbital eccentricity $e$
\begin{equation}
F(e)= \frac{1+\frac{73}{24}e^2 + \frac{37}{96}e^4}{(1-e^2)^{7/2}}.
\end{equation}
From Eq. \ref{eq:tgw} we can derive that at $M_{\bullet\bullet}=10^9$ and $T_{\rm orb}=2$ yr the GW-driven orbital decay timescales are $\sim 1.2\times 10^5$ and $\sim 4.5\times 10^4$ yr for $q_M=0.1$ and $0.5$ at $e=0.5$, respectively.
The orbital decay would not affect the periodicity of BBHs in this work significantly. 

\subsection{An optimized way of identifying BBH candidates}

Currently, there are mainly two efficient ways of searching for BBH candidates, i.e., by selecting periodic light curves \citep[e.g.,][]{DOrazio2015,Graham2015,Charisi2016,Liu2019,Chen2022} or based on asymmetric/double-peaked BEL profiles \citep[e.g.,][]{Eracleous1994,Gaskell1996,Tsalmantza2011,Eracleous2012,Popovic2012,Decarli2013,Ju2013,Shen2013,Liu2014,Guo2019}. However, both the dynamical modelling of BBHs \citep[e.g.,][]{DOrazio2015,Jiang2022} and BEL profile modelling \citep[e.g.,][]{Shen2010,Nguyen2019,Nguyen2020} are hard to be distinguished from the single BH case due to a set of model degeneracy \citep[see reviews by][]{WL2020,DC2023}. 
Limited by photometric and spectroscopic observations, current efforts on the selection and identification of BBH candidates have mainly focused on the continuum light curves or BEL profiles separately. In \cite{Ji2021a,Ji2021b}, we have analyzed the detailed response of BLR clouds and BEL profiles to the central BBHs that co-rotating in circular orbits, by focusing on the cases that the BBH orbital plane are co-planar \citep{Ji2021a} or misaligned \citep{Ji2021b} with the middle plane of the circumbinary BLR. In this paper, we investigate the coherent variation of BBH continuum light curves and BEL profiles for BBHs orbiting in different eccentricities with simplified orbital orientation, which indicates that the joint analyses of periodic light curves and multi-epoch observed BELs could lead to successful identification of BBH candidates.

As analyzed in this work, the amplitudes of periodic continuum light curves and BELs are correlated with each other (see Figs. \ref{Fig:secBH_both_conti_BEL_amplitude} and \ref{Fig:secBH_both_conti_BEL_diffiobs}).
Because of the momentum conservation, the active primary BH could cause non-sinosoidal continuum light curves and decreased $A_{\rm conti}$ (Figs. \ref{Fig:conti_light_curve}-\ref{Fig:conti_light_curve_2active}). The BEL profiles at different phases in one period not only have significant difference in shapes and fluxes (Figs. \ref{Fig:circ_orbits}-\ref{Fig:secBH_both_iobs30}) as caused by orbital eccentricity and argument of periapsis of BBHs at certain ($q_M,q_L$), but also have the flux variation amplitudes decrease with increasing mass ratio and luminosity contribution by the primary BH (Figs. \ref{Fig:secBH_both_conti_BEL_amplitude} and \ref{Fig:secBH_both_conti_BEL_diffiobs}). When taking into account the inclination angle, both the amplitudes of continuum light curves and BEL profiles increase with increasing inclination angles. This means that modelling the continuum light curves and multi-epoch observed BEL profiles together would improve the probability of identifying BBH candidates, and the BBH scenario can be confirmed if a clear coherent variation of the BELs with the continuum can be found.

On the other hand, the on-going \citep[e.g., WFST,][]{WangT2023} and future photometric surveys \citep[e.g., the Rubin and Sitian project,][]{Ivezic2019, Liu2021} would find more BBH+cBLR systems occupied at much wider parameter space compared with those currently obtained by CRTS, Pan-STARRS, PTF, and ZTF projects \citep[e.g.][]{Graham2015,Charisi2016,Liu2019,Chen2022}. And there will be more spectroscopic data obtained from DESI \citep{DESIsurvey2016}, the Nancy Grace Roman Space Telescope \citep{Spergel2015}, and SDSS-V \citep{Kollmeier2017} surveys. This makes the simultaneous modelling of periodic light curves and BEL profiles being possible for many BBH candidates.

\section{Conclusion}

In this paper, we investigate the response of the line emission from BLR clouds to the continuum emission from central BBH systems co-rotating in elliptical orbits, by focusing on the evolution trends of both observed light curves and BEL profiles. For QSOs with the central BBHs surrounded by a circumbinary BLR (BBH+cBLR) system, we investigate two cases of BH activities: 1) only the secondary BH is active ($q_L=1:0$), and 2) both the two BHs are active with equal luminosity ($q_L=1:1$), with simplified orbital orientation, i.e., $\Omega=0^\circ$. 

For BBHs co-rotating in elliptical orbits instead of circular ones, the Doppler boosting effect caused by their orbital eccentricity and argument of periapsis could affect both the shape of continuum light curve, e.g., sharply (slowly) increased but slowly (sharply) decreased trends in a period, and two kinds of periodicities in BEL profile variations in the Doppler Boosting hypothesis.
On the other hand, BBHs in different active behaviours can result quite different amplitudes of light curves and BEL profiles. At $q_L=1:0$, both the variation amplitudes of continuum light curves and BELs increase with increasing eccentricity but decrease with increasing mass ratio of BBHs. While for $q_L=1:1$, the activity of the primary BH can cause systematically decreased amplitudes of both continuum light curves and BELs due to its enhanced ionization to cBLR clouds in the Doppler weakened area that modulated by the secondary BH. Currently, we only consider BHs accrete with stable accretion rate, in future, we will further explore how the BBH systems vary with different fluctuations in accretion rates, such as the damped random walk variation.

The confirmation of BBH candidates becomes more complex when considering the elliptically co-rotated orbits and accretion activities of BBHs than those cases with simply assumed circular orbits or single BH activity. With the theoretical understanding on BBH systems at varied mass ratios, luminosity ratios, orbital eccentricities, and BH accretion activities, the incoming huge amount of photometric (e.g., by Rubin, Sitian, and WFST) and multi-spectroscopic (e.g., by DESI, Roman space telescope, and SDSS-V) observations would search for BBH candidates with more complete parameter spaces as mentioned in this work, based on which we can constrain the merger rate of massive black hole binaries.

\begin{acknowledgements}
We would like to thank the anonymous referee for the suggestions that helped us to improve this paper.
This work is supported by the National Key program for Science and Technology Research and Development (grant Nos. 2020YFC2201400 and 2022YFC2205201), the Beijing Municipal Natural Science Foundation (No. 1242032), the National Natural Science Foundation of China (NSFC) (grant Nos. 11903046, 12273050, 11991052, 11988101, and 11933004), and the Strategic Priority Program of the Chinese Academy of Sciences (Grant No. XDB23040100).
JG acknowledges support from the Youth Innovation Promotion Association of the Chinese Academy of Sciences (No. 2022056) and the science research grants from the China Manned Space Project, and JFL acknowledges support from the New Cornerstone Science Foundation through the New Cornerstone Investigator Program and the XPLORER PRIZE.
\end{acknowledgements}




\begin{appendix}

\section{Stability of the accretion disks for BBH systems} \label{accretion_disk_stab}
\begin{figure*}
\centering
\includegraphics[angle=0.0,scale=0.95,origin=lb]{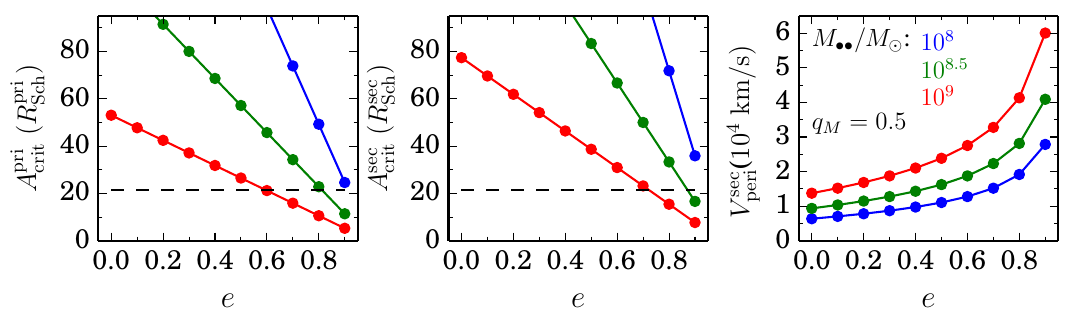}
\caption{The radii of Roche lobes of BBHs and rotating velocity of the secondary BH at the orbital pericenter as a function of orbital eccentricity of the two BHs. The left and middle panels show the radii of Roche lobes ($R_{\rm crit}^{\rm pri}$ and $R_{\rm crit}^{\rm sec}$ in unit of $R_{\rm Sch}^{\rm pri}$ and $R_{\rm Sch}^{\rm sec}$, respectively) in units of the Schwarzschild radii for the primary ($R_{\rm Sch}^{\rm pri}$) and secondary ($R_{\rm Sch}^{\rm sec}$) BHs, respectively. The typical radius of the UV emission ($r_{\rm UV}$) in the accretion disk for a $M_{\bullet}=10^{8.6}M_{\sun}$ is shown as a horizontal dashed line for comparison. The right panel presents the rotating velocity of the secondary BH at the pericenter ($V_{\rm peri}^{\rm sec}$), which is the maximum velocity in each period. In each panel, we show results of $M_{\bullet\bullet}=10^8$ (blue color), $10^{8.5}$ (green color), and $10^9M_{\sun}$ (red color) with $T_{\rm orb}=2$ yr and $q_M=0.5$. Considering that the UV emission might disappear when $R_{\rm crit}\lesssim r_{\rm UV}$, current model configuration only focus on those systems with $R_{\rm crit}\gtrsim r_{\rm UV}$, i.e., $e\lesssim 0.6$ for $M_{\bullet\bullet}=10^9M_{\sun}$ as a case study.}
\label{Fig:BBH_dist_vel}
\end{figure*}

For the active BH in the BBH system, we simply assume that the BH accretion disk is steady inside the Roche lobe. When studying the photoionization process of UV photons that radiated from the accretion disk to BLR clouds, the minimum radius of the accretion disk should make sure the quantity of radiated UV photons is enough for ionizing BLR clouds and result matchable continuum and BEL fluxes as observed from those single BH cases, i.e., the radius of Roche lobe $R_{\rm crit}$ is equal to or larger than the half light radius of the accretion disk at UV band ($r_{\rm UV}$). Only with $R_{\rm crit}\gtrsim r_{\rm UV}$ the BEL profiles could be observed and encode the dynamics of the BBH system. For BBHs systems \citep{Eggleton1983}, the radius of the Roche lobe for the BH could be appoximated within $1\%$ of its value with the formula
\begin{equation}
R_{\rm crit} = \frac{0.49q_M^{2/3}}{0.6q_M^{2/3}+\ln(1+q_M^{1/3})}.
\end{equation}

At the pericenter, the minimum radius of the Roche lobe for the secondary BH is $R_{\rm crit}^{\rm min}=R_{\rm crit}(1-e)$.
If the radius of the accretion disk that emitting UV photons is smaller than the radius of the Roche lobe, the BBH+cBLR system would be reasonably established for reflecting the BBH features by reverberation mapping procedures. For a BH with $M_{\bullet}=10^{8.6}M_{\odot}$, the typical radius of the UV continuum emission at 1350\AA\ in the accretion disk is $r_{\rm UV}^{1350\AA}=21.4~ R_{\rm Sch}$ for the QSO SDSS J1339+1310 \citep{Shalyapin2021}. If estimating $r_{\rm UV}$ at 2500 \AA\ with the empirical correlation derived from 14 QSOs \citep[][]{Morgan2018}, the BH masses ranging from $10^8$ to $10^9 M_{\sun}$ correspond to $r_{\rm UV}^{2500\AA}\sim 20$ to $30~R_{\rm Sch}$. Given a fixed BH mass and accretion rate, $r_{\rm UV}^{1350\AA}$ is smaller than $r_{\rm UV}^{2500\AA}$. Fig. \ref{Fig:BBH_dist_vel} shows $R_{\rm crit}$ as a function of $e$, from which we can see that when studying a BBH system with $M_{\bullet\bullet}=10^9M_{\sun}$ and $T_{\rm orb}=2$ yr, focusing on those BBHs with orbital eccentricities $e\lesssim 0.6$ would be reasonable (left and middle panels) if using $r_{\rm UV}^{1350\AA}$ of SDSS J1339+1310 as a criteria, in which the rotational velocity of the secondary BH that ranging from $\sim 10^4$ km/s to $\sim 3\times 10^4$ km/s (right panel) could results significant Doppler boosting effect to both photoionization of BLR clouds and flux variation received by observers.

\section{Dynamics of the BBH system} \label{BBH_dynamics}

Assuming an BBH system with a orbital period of $T_{\rm orb}=2$ years, which is roughly the median intrinsic period of BBH candidates in the sample of \cite{Charisi2016}, the semi-major axis of the two black holes with total mass $M_{\bullet\bullet}=10^9 M_{\odot}$ is
\begin{equation}\label{EQ:a_BBH}
a_{\rm BBH} = 7.7\times 10^{-3}\left(\frac{M_{\bullet\bullet}}{10^9 M_{\odot}}\right)^{1/3} 
\left(\frac{T_{\rm orb}}{2\, \rm yr}\right)^{2/3} \mbox{pc},
\end{equation}

In the case of elliptical orbits \citep[e.g.,][]{Valtonen2008, Charisi2022}, the velocity of the secondary BH depends on its location at different position angle. 
Given the secondary BH rotating in elliptical orbit with eccentricity $e$ and mass ratio $q_M$ in the X-Y plane (Fig. \ref{Fig:ecc_orbits}), the radius of the secondary BH is
\begin{equation}
r_{\rm sec} = \frac{a_{\rm BBH}(1-e^2)}{(1+q_M)(1+e\cos\phi)},
\end{equation}
and the velocity components $V_{\phi}$ and $V_r$ of the secondary BH are 
\begin{equation}
V_{\phi} = r_{\rm sec}\frac{d\phi}{dt} = \sqrt{\frac{GM_{\bullet\bullet}}{a_{\rm BBH}(1-e^2)}}\frac{1+e\cos\phi}{1+q_M},
\end{equation}
\begin{equation}
V_r = \frac{dr_{\rm sec}}{dt} = \sqrt{\frac{GM_{\bullet\bullet}}{a_{\rm BBH}(1-e^2)}}\frac{e\sin\phi}{1+q_M}.
\end{equation}
where $G$ is the acceleration of gravity, $\phi$ is the true anomaly of the secondary BH. The position and velocity of the primary BH can then be derived by momentum conservation.

\section{The simplified BLR model} \label{BLR_model}

For a BBH system with a cBLR (BBH+cBLR), we simply assume its BLR size has the same relation with the luminosity of the central ionization source as that of single BH systems \citep[][]{Kaspi2000, Kaspi2007, MJ2002, Peterson2004}, although it could be different. The BLR size of the BBH system can then be derived by following the empirical relationship between the BLR size and optical luminosity \citep[e.g.,][]{Lu2016}. For convenience of model constructing, we convert the optical luminosity to the Eddington ratio ($\lambda_{\rm Edd}$) and $M_{\bullet\bullet}$ by assuming $\lambda_{\rm Edd} = 0.1$ for the BBHs:
\begin{equation}
R_{\rm BLR} = 0.127 \times (\lambda_{\rm Edd}/0.1)^{0.54} (M_{\bullet\bullet}/10^9 M_\odot)^{0.54}~ {\rm pc},
\end{equation}
The typical BLR size is $\sim 8$ times larger than the separation of the BBHs, significantly far away from the central two BHs, and can be taken as a cBLR affected little by the central BBH dynamics. Therefore, in this work, we consider the configuration of BBHs surrouding with a cBLR, for which we take a shifted $\Gamma-$distribution \citep[see details in][]{Pancoast2014a} as the radial emissivity distribution of BLR clouds:
\begin{equation}
R_{\rm ga} = R_{\rm S} + F R_{\rm BLR} +g(1-F) \beta^{2} R_{\rm BLR},
\end{equation}
where $R_{\rm S} $ is the Schwarzschild radius, $R_{\rm BLR}$ is the mean BLR radius, F = $R_{\rm min}/R_{\rm BLR}$ is the fractional inner BLR radius ($R_{\rm min}$), $\beta$ is the shape parameter, and $g = p(x|1/\beta^{2},1)$ is drawn randomly from a Gamma distribution. 

As to the angular distribution of the BLR clouds, for a BLR with an half opening angle $\theta_{\rm o}$ (Fig. \ref{Fig:ecc_orbits}), we assume 
\begin{equation}         
\theta = \arccos \left[ \cos \theta_{\rm o} +(1-\cos \theta_{\rm o})   U \right],
\label{eq:theta}
\end{equation}
where $\theta$ represents the angle between the orbital angular momentum of a cloud and the normal to the BLR middle plane, and $U$ is a random number ranging from 0 to 1 and is set to describe the randomly distributed clouds but constrained with $\theta \le \theta_{\rm o}$ \citep[see][for details]{Pancoast2014a}.

\section{Response of BLR clouds to the central source} \label{BLR_response}
For the response of BLR clouds to the central BBH systems co-rotating in elliptical orbits, the equations are similar to that with BBHs co-rotating in circular orbits as introduced in our previous work \citep{Ji2021a}, except for differential rotating time in each orbital phase and the variation of projected major axis of the elliptical orbit.

For the continuum emission dominated by the active secondary black hole, the time in the observer's frame ($t_{\rm obs}$) is determined by the the intrinsic time ($t_{\rm in}$) in the BBH mass center rest frame (hereafter BBH frame) and the cosmological time dilation, i.e., 
\begin{eqnarray}
t_{\rm obs}-t_{\rm obs,0} & = & \left[(t_{\rm in} -t_{\rm in,0})+ \frac{\overrightarrow{\rm \boldsymbol{OS} }\cdot {\rm \boldsymbol{\hat{l}} }}{c}\right](1+z),  
\end{eqnarray}
where $t_{\rm obs,0}$ is the starting time for the observation, $t_{\rm in}$ and $t_{\rm in,0}$ are the times in the BBH frame corresponding to $t_{\rm obs}$ and $t_{\rm obs,0}$, respectively. $\overrightarrow{\rm \boldsymbol{OS} }$ is the position vector of the secondary black hole at the time $t_{\rm in}$, and $\hat{\boldsymbol{l}}$ is the unit vector along the line of sight. For convenience, we set $t_{\rm obs,0}=0$ and $t_{\rm in,0}=0$, when the secondary BH is located at the apocenter of the elliptical orbit.

For photons emitted from a cloud C in the BLR located at $(x_{0},y_{0},z_{0})$, the observation time $t_{\rm obs}$ is related to the intrinsic time $t_{\rm in}^{'}$ in the BBH frame when the secondary BH is located at $S^{\prime}$, i.e.,
\begin{eqnarray}
& t_{\rm obs}-t_{\rm obs,0}  =  \left[ (t_{\rm in}^{'} -t_{\rm in,0}) + \frac{\overrightarrow{\rm \boldsymbol{OC} }\cdot \rm {\rm \boldsymbol{\hat{l}} }+ |\overrightarrow{\boldsymbol{S^{\prime}C} }|}{c} \right] (1+z), 
\end{eqnarray}
where $\overrightarrow{\rm \boldsymbol{OC}}$ is the position vector the cloud C, $\overrightarrow{\boldsymbol{S^{\prime}C}}$ is the distance between the secondary BH and the cloud C. At a specific moment of observation $t_{\rm obs}$, we can derive the emission time difference between the continuum observed by the observer and that received by the BLR cloud is
\begin{eqnarray}
\tau = t_{\rm in} -t_{\rm in}^{'}
=\left( \overrightarrow{\rm \boldsymbol{OC} }\cdot \hat{\rm \boldsymbol{l} } +|\overrightarrow{\boldsymbol{S^{\prime}C} }| -
\overrightarrow{\rm \boldsymbol{OS} }\cdot \hat{\rm \boldsymbol{l} }\right) /c,
\end{eqnarray}
The above equations are the same for the BBH systems co-rotating both in circular and elliptical orbits.

For the modelling of the BBH+cBLR system under the Doppler boosting scenario, we assume the accretion rates are constant \citep[e.g.,][]{DOrazio2015, Duffell2020}, whenever only the secondary BH is active or both BHs are active. Five factors are taken into account to derive the finally observed BEL profiles \citep[see details in][]{Ji2021a, Ji2021b}:
\begin{itemize}
\item The position variation of the two BH components;
\item The time-dependent Doppler boosting effect on the ionizing flux emitted from the primary/secondary BH accretion disks and recieved by the BLR clouds;
\item The reverberation process of the BLR clouds to the central source(s);
\item The gravitational redshift of the emission from BLR clouds \citep[][]{Tremaine2014};
\item The Doppler blue-/red-shifts of photons from the BLR clouds to the observer.
\end{itemize}

By assuming that the intrinsic continuum flux emitted from the disk around the primary/secondary BH can be described by a power law $F_{\nu, \rm e}(\nu; t) \propto \nu^{\alpha}$, in which we assume the spectral index $\alpha=-2$ \citep[e.g.,][]{DOrazio2015, Song2020, Song2021, Ji2021a, Ji2021b}, flux emitted from each BLR cloud ${\rm C}_i$ can be derived. We can then obtain the BLR radiation (see \citet[][]{Ji2021a} for more details):

\begin{eqnarray}
L(v,t_{\rm obs})
& \propto & \left. \sum_{i=1}^{N_{\rm tot}} F_{\nu,\rm e}(\nu; t_{\rm in}^{'}) D^{3-\alpha}_{{\rm C}_i2}\frac{|\vec{r}_{{\rm C}_i}|^{2}}{|\vec{r}_{{\rm C}_i} - \vec{r}_{\rm BH}|^2}  \right|_{v= { v^{\rm tot}_{{\rm C}_i}}}.
\label{eq:f_BLR}
\end{eqnarray}
where $N_{\rm tot}$ is the total number of BLR clouds, $D_{{\rm C}_i2}^{3-\alpha}$ represents the Doppler boosted/weakened ionizing flux $F_{\nu,\rm e}(\nu; t_{\rm in}^{'})$ emitted at $t_{\rm in}^{'}$ and received by the BLR cloud at $t_{\rm obs}$ due to the relativistically orbital motion of the primary/secondary BH, and the term $|\vec{r}_{{\rm C}_i}|^2/|\vec{r}_{{\rm C}_i} -\vec{r}_{\rm BH}(t'_{\rm in}) |^2$ reflects the continuum flux variation due to the moving primary/secondary BH, $v^{\rm tot}_{{\rm C}_i} \simeq v^{\rm D}_{{\rm C}_i} + v^{\rm g}_{{\rm C}_i} $ represents the summation of the Doppler redshift/blueshift  $v^{\rm D}_{{\rm C}_i}=\left\{\left[ \gamma_{{\rm C}_i{\rm obs}}(1-\vec{\beta}_{{\rm C}_i{\rm obs}}\cdot \boldsymbol{\hat{l}})\right]^{-1}-1\right\} c$  and the gravitational redshift of photons emitted from an individual BLR cloud received by the distant observer $v^{\rm g}_{{\rm C}_i} =  GM_{\bullet\bullet}/(r_{\rm C_{i}}c) $  \citep[][]{Tremaine2014}, $\vec{\beta}_{{\rm C}_i{\rm obs}}=\vec{v}_{{\rm C}_i{\rm obs}}/c$ is the observed velocity of the cloud C$_i$ in unit of the light speed, $\gamma_{{\rm C}_i{\rm obs}}=1/\sqrt{1-|\vec{\beta}_{{\rm C}_i{\rm obs}}|^2}$, and $\vec{r_{\rm C_{i}}}$ is the distance vector of the BLR cloud to the mass center of the BBH system.  In each model, the line emissivity from each cloud is assumed to be proportional to the flux received by the cloud.

\end{appendix}

\end{document}